# Giant nonlinearity in Zero-Gap Semiconductor Superlattices

*Mário G. Silveirinha*[(1, 2)*] *and Nader Engheta*[(1)]

*(1) University of Pennsylvania, Department of Electrical and Systems Engineering, Philadelphia, PA, U.S.A., engheta@ee.upenn.edu*
*(2) University of Coimbra, Department of Electrical Engineering – Instituto de Telecomunicações, Portugal, mario.silveirinha@co.it.pt*

**Abstract**

Using an effective medium approach, we describe how by combining mercury-cadmium-telluride semiconductor alloys with band gap energies of opposite signs it may be possible to design a superlattice where the electrons have isotropic zero-effective mass and a single valley linear energy-momentum dispersion. We demonstrate that because of the zero-mass property, the superlattice may have a strong nonlinear response under the action of an external electromagnetic field.

---

[*] To whom correspondence should be addressed: E-mail: mario.silveirinha@co.it.pt

# I. Introduction

Semiconductor superlattices were originally proposed in a seminal work by Esaki and Tsu in late sixties of the last century [1]. The superlattice idea consists of creating, either by doping or by changing periodically the material composition or by other means, nanoscopic periodic features that can influence the propagation of electron waves. Perhaps the forefront application of the superlattice idea is band-engineering [2] such that by controlling the nanoscopic features of the superlattice it is possible to tailor the electronic structure of the energy stationary states, and as a consequence the transport and optical properties of the structure. Band engineering has played a key role in the development of semiconductor based electronics and light sources [3-4].

Several recent works have highlighted that superlattices may be regarded as the semiconductor counterpart of the modern concept of electromagnetic metamaterials [5-9]. In particular, a direct analogy between the electron wave propagation in III-V and II-VI semiconductor superlattices and the light wave propagation in a metamaterial becomes manifest if one describes the propagation of electron waves using the envelope function approximation developed by Bastard [5-11]. We explored this analogy in Ref. [7] to demonstrate the possibility of a perfect lens for electron waves, and in Ref. [8] to propose a novel artificial electronic material characterized by an extreme anisotropy of the electron effective mass, such that the mass is ideally zero along some preferred direction of motion (the direction of growth of the stratified superlattice) and infinite for perpendicular directions. Here, building on our previous works [7, 8, 9], we investigate the transport properties and nonlinear response of semiconductor superlattices with an *isotropic* zero-effective mass and linear energy dispersion.

Zero-gap semiconductor superlattices and related structures [8, 12-21] have recently elicited significant attention because they may behave as "artificial" graphene-like system that mimic Dirac-type electron transport [22, 23] (see also Refs. [11, 24-26] for earlier studies related to zero-gap semiconductors). These materials may open new routes for ultrahigh-mobility electrons with linearly dispersing energy bands. Here, we explore the possibility of designing a zero-gap material with a single Dirac cone based on Mercury-Cadmium-Telluride (HgCdTe) superlattices. Our theoretical framework is based on the envelope function approximation. It is demonstrated that the superlattice may be regarded as an "effective medium" characterized by a dispersive (energy dependent) effective mass and an effective potential that allow characterizing the stationary states of the material. We prove that the effective parameters can be written in terms of the constituents of the superlattice through mixing formulas that are analogous to the classical "Clausius-Mossotti" formulas well-known in electromagnetics. We validate our theory through exact band structure calculations based on the plane wave method.

It is well known that semiconductors heterostructures may give large optical nonlinearities. For example, coupled quantum well semiconductors enable giant nonlinear optical responses for second and third harmonic generation associated with intersubband transitions in the infrared [27-30]. More recently, it has been shown that the electrical response of graphene can be highly nonlinear [31-33] due to the zero mass property. Here, motivated by these earlier studies, we investigate the nonlinear response of the superlattice under the excitation of an external electromagnetic field. We prove that in case of a zero gap the nonlinear electrical response for the third harmonic may be extremely strong at terahertz frequencies, several orders of magnitude larger than in natural materials.

## II. Theoretical Framework

We consider a superlattice formed by two bulk semiconductor materials (Fig. 1). The pertinent materials are assumed to be lattice matched, and possible effects of strain [34] are disregarded. Within the envelope function ($\Psi$) approximation [10, 11], the electronic structure of binary compounds with a zincblende-type structure can be determined based on a potential $V = V(E)$ and a dispersive (energy dependent) effective mass parameter $m = m(E)$, which characterize the effective Hamiltonian of a single electron as detailed in Refs. [8, 9].

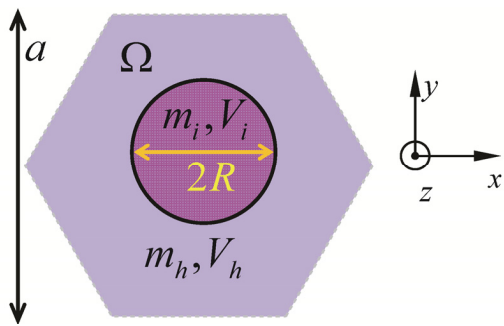

Fig. 1. Representative geometry of the unit (Wigner-Seitz) cell ($\Omega$) of a triangular superlattice. The host semiconductor compound is characterized by the parameters $V_h$ and $m_h$, and the "inclusion" by the parameters $V_i$ and $m_i$. The lattice constant is $a$.

For narrow gap binary compounds of the groups III-V and II-VI the dispersive mass may be assumed to be a linear function of the energy, such that for each semiconductor compound it is of the form [9]

$$m = \frac{1}{2v_P^2}(E - E_v), \tag{1}$$

with $E_v = E_{\Gamma_8}$ being the energy level associated with the edge of the light-hole band. The effect of the split-off band is negligible in the linear mass dispersion approximation. The velocity $v_P = \sqrt{\varepsilon_P/(3m_0)}$ is written in terms of the Kane's $P$ parameter using $\varepsilon_P = 2P^2 m_0/\hbar^2$, with $m_0$ being the free electron rest mass. On the other hand, for bulk semiconductors the effective potential may be assumed energy independent, and satisfies

$$V_{ef}(E) = E_c, \qquad (2)$$

with $E_c = E_{\Gamma_6}$ the conduction band edge energy.

We use the envelope function formalism to calculate the electronic structure of a semiconductor superlattice. The stationary energy states can be determined by solving a Schrödinger-type equation of the form [8, 9, 10]:

$$-\frac{\hbar^2}{2}\nabla\cdot\left(\frac{1}{m}\nabla\Psi\right) + V\Psi = E\Psi. \qquad (3)$$

Evidently, in case of a superlattice (Fig. 1), the material parameters must be regarded functions of position, so that $m = m(\mathbf{r}, E)$ and $V = V(\mathbf{r}, E)$. Note that because $m$ depends on the position, $\nabla\cdot\left(\frac{1}{m}\nabla\Psi\right) \neq \frac{1}{m}\nabla^2\Psi$. The adopted form $\nabla\cdot\left(\frac{1}{m}\nabla\Psi\right)$ is consistent with the analogy with electromagnetic metamaterials discussed in Ref. [9], and also with the generalized Ben Daniel-Duke boundary conditions [10, 34, 35], because the structure of Eq. (3) implies that $\Psi$ and $\frac{1}{m}\frac{\partial\Psi}{\partial n}$ are continuous at an abrupt interface between two materials. The Ben Daniel-Duke boundary conditions ensure the conservation of the probability current at an interface.

We consider a "two-phase" superlattice with the generic geometry of Fig. 1 such that $m = m(\mathbf{r}, E)$ and $V = V(\mathbf{r}, E)$ are piecewise continuous with values $V = V_h$ and $m = m_h$ in the host material and $V = V_i$ and $m = m_i$ in the inclusions. Moreover, for simplicity it is assumed that the Kane velocity $v_P = \sqrt{\varepsilon_P / (3m_0)}$ is the same in the host and inclusion regions, which is equivalent to saying that Kane's $P$ parameter has the same value in the two pertinent bulk semiconductor compounds (this is a good approximation for the pair of materials considered in the numerical example). In such a case one may write:

$$m_h = \frac{1}{2v_P^2}(E - E_{v,h}) \quad ; \quad m_i = \frac{1}{2v_P^2}(E - E_{v,i}) \qquad (4)$$

For a given geometry and material parameters, the electronic structure of the superlattice can be calculated by numerically solving Eq. (3), and this has been done in many previous works (e.g. [11]). In Appendix A, we describe a formulation that enables calculating the electronic band structure (Bloch modes of the generalized Schrödinger equation) of 2D-superlattices under the considered theoretical framework.

However, our objective here is more ambitious, and we want to prove that analogous to electromagnetic metamaterials, it is possible to regard the superlattice itself as an effective medium, characterized by effective parameters $V_{ef}$ and $m_{ef}$, which depend on the "nanoscopic" parameters $m = m(\mathbf{r}, E)$ and $V = V(\mathbf{r}, E)$. Obviously, such a description involves a second level of homogenization, not on the scale of the atomic unit cell of each bulk semiconductor, but on the scale of the unit cell of the superlattice. This is the topic of the next section.

# III. Effective Medium Theory

Our objective is to determine an effective Hamiltonian $\hat{H}_{ef}$ operator of the form,

$$\hat{H}_{ef} = -\frac{\hbar^2}{2}\nabla \cdot \frac{1}{m_{ef}}\nabla + V_{ef}. \tag{5}$$

that describes the dynamics of "macroscopic" initial electronic states in the superlattice as well as the energy stationary states, as detailed in Refs. [9, 36]. In this approach, the superlattice is seen as a continuous medium with no granularity.

The effective potential $V_{ef}$ may be estimated simply as the spatial averaged value of $V = V(\mathbf{r}, E)$ so that:

$$V_{ef} = V_h(1 - f_V) + V_i f_V \tag{6}$$

where $V = V_h$ in the host material, $V = V_i$ in the inclusions, and $f_V$ is the volume fraction of the inclusions. This is justified by the fact that for stationary states with a quasi-momentum $\mathbf{k} \approx 0$, the solutions of Eq. (3) may vary relatively slowly over the scale of the unit cell of the superlattice ($\Psi \approx const.$), and hence $\langle V\Psi \rangle \approx \langle V \rangle \langle \Psi \rangle$ where $\langle \ \rangle$ denotes the spatial-averaging operator [9, 36].

Rather than attempting a formal derivation of $m_{ef}$, we appeal to the analogies between electronics and electromagnetics [9]. The details are given in Appendix B, where it is shown that the dispersive effective mass $m_{ef}$ for a superlattice formed by an array of "spherical" inclusions (e.g. quantum dots) with dispersive mass $m_i$ embedded in a host with dispersive mass $m_h$ is given by,

$$m_{ef} = m_h \frac{(1 - f_V)m_h + (2 + f_V)m_i}{(1 + 2f_V)m_h + 2(1 - f_V)m_i} \qquad \text{(spherical inclusions)} \tag{7}$$

where $f_V$ is the volume fraction of the inclusions. The inclusions are assumed to be packed in a highly symmetric lattice (e.g. with a f.c.c., b.c.c, or s.c. structure). It should be mentioned that our model of spherical inclusions is only an idealized mathematical model, since each bulk material has an intrinsic granularity (determined by the atomic lattice constant $a_s$) that in practice is only one or two orders of magnitude smaller than the typical period of a superlattice. As mentioned previously, we neglect any possible defects at the interfaces between the host and the inclusions, either related to a possible mismatch of the lattice constants or to other factors such as disorder.

It is also useful to find the effective parameters for a superlattice formed by cylindrical inclusions with circular cross-section and oriented along the $z$-direction. Restricting our attention to stationary states such that $\partial/\partial z = 0$ it is simple to show that the effective mass in such a scenario is given by (Appendix B):

$$m_{ef} = m_h \frac{(1-f_V)m_h + (1+f_V)m_i}{(1+f_V)m_h + (1-f_V)m_i} \qquad \text{(cylindrical inclusions)} \qquad (8)$$

It is assumed that the inclusions are packed in a lattice with high symmetry (e.g. square lattice or a triangular lattice). The effective potential is still given by Eq. (6).

Even though the above derivations of $V_{ef}$ and $m_{ef}$ may look a bit heuristic, in the next section the model is validated against full wave simulations based on the exact numerical solution of Eq. (3). Within the validity of the effective medium theory [9, 36], the stationary Bloch states of the superlattice satisfy:

$$\frac{1}{2}\frac{\hbar^2 k^2}{m_{ef}(E)} + V_{ef} = E, \qquad (9)$$

where $m_{ef}(E)$ is given by Eq. (8) in case of cylindrical inclusions and $V_{ef}$ by Eq. (6). Note that in general $m_{ef} = m_{ef}(E)$ because $m_i$ and $m_h$ are energy dependent [Eq. (4)].

## IV. HgCdTe Superlattices with Zero-Gap

Here, we want to investigate the possibility of realizing single-valley graphene-like systems, with a zero-gap and linear energy dispersion, based on HgCdTe superlattices. Because the exact numerical solution of Eq. (3) requires significant numerical effort, first we concentrate on the case wherein the inclusions are shaped as cylinders (wires) so that the problem is effectively two-dimensional. The case of dot-type inclusions is discussed in Sect. V. We use the effective medium theory developed previously to design the zero-gap superlattice.

From the theory of Refs. [8, 9], it is expected that the edge of the conduction band of the superlattice occurs at the energy level $E = V_{ef}$ [see Eq. (6)], whereas the edge of the valence band occurs at the energy level such that $m_{ef}(E) = 0$ [see Eq. (8)]. To have a superlattice with a zero-gap, the edge of the conduction band must be coincident with the edge of the valence band. Thus, the zero-gap condition is simply

$$m_{ef}(E = V_{ef}) = 0. \tag{10}$$

Substituting Eqs. (4), (6), (8) into Eq. (10), it is found that the zero-gap condition requires that the volume fraction of the inclusions is chosen equal to:

$$f_{V0} = \frac{E_{v,h} + E_{v,i} - 2V_h}{E_{v,h} - E_{v,i} - 2(V_h - V_i)}, \qquad \text{(for cylindrical inclusions)}. \tag{11}$$

The corresponding effective potential of the superlattice is given by:

-9-

$$V_{ef0} = \frac{(E_{v,h} + E_{v,i})V_i - 2E_{v,i}V_h}{E_{v,h} - E_{v,i} - 2(V_h - V_i)}. \tag{12}$$

It should be mentioned that because of the granularity of the superlattice constituents $f_V$ cannot vary continuously, but rather varies in discrete steps. Here, we will not be concerned with these difficulties and regard $f_V$ as a continuous variable. Evidently, this approximation is better for large values of $a/a_s$, so that the atomic building blocks of the superlattice are much smaller than the period of the superlattice.

To illustrate the application of the derived formulas and validate our effective medium approach, we consider a superlattice in which the host material is the ternary compound Hg$_{0.75}$Cd$_{0.25}$Te and the inclusions are made of HgTe. These semiconductor alloys are (nearly) lattice matched, with an atomic lattice constant $a_s = 0.65 nm$, and have $v_{P,h} \approx v_{P,i} \equiv v_P = 1.06 \times 10^6 m/s$ [37]. We suppose that the inclusions are arranged in a triangular lattice, defined by the primitive vectors $\mathbf{a}_1 = a(\sqrt{3}/2, 1/2)$ and $\mathbf{a}_2 = a(0,1)$, where the period is taken equal to $a = 12a_s$. The band gap energy $E_g = E_g(x)$ of the ternary compound Hg$_x$Cd$_{1-x}$Te is calculated with Hansen's formula at zero temperature [37], where $x$ represents the mole fraction. Thus, the potential in each material can be written as:

$$V_h \equiv E_{c,h} = E_{v,h} + E_{g,x=0.25}, \qquad V_i \equiv E_{c,i} = E_{v,i} + E_{g,x=0} \tag{13}$$

On the other hand, the valence band offset $\Lambda(x) = E_{v,x=0} - E_{v,x}$ between Hg$_{1-x}$Cd$_x$Te and HgTe, can be estimated by linearly interpolating the values of $\Lambda$ for $x=0$ [$\Lambda = 0$] and for $x=1$ [$\Lambda = 0.35 eV$][38]. This yields $\Lambda(x) = 0.35x \, [eV]$, so that:

$$E_{v,h} = E_{v,i} - \Lambda_{x=0.25} . \tag{14}$$

The value of $E_{v,i}$ can be arbitrarily chosen, and fixes the reference energy level. In this work, $E_{v,i}$ is chosen in such a manner that $V_{ef0}$ defined as in Eq. (12) vanishes: $V_{ef0} = 0$. The graphic in the lower right-hand side corner of Fig. 2 shows the electronic structure of the bulk semiconductor alloys. The blue lines represent the conduction (solid curve) and valence band (dashed curve) of $Hg_{0.75}Cd_{0.25}Te$, respectively, whereas the black lines are associated with HgTe. As seen, the band structure of HgTe is inverted, so that the valence band lies above the conduction band, and the band gap energy is negative.

We calculated the energy band structure of the superlattice as a function of the normalized volume fraction of the HgTe inclusions. The critical volume fraction for which there is a zero-gap is $f_{V0} = 0.247$ [Eq. (11)]. In Fig. 2 the calculated energy dispersions are depicted for different values of the normalized volume fraction $f_V / f_{V0}$ (indicated as an inset in the graphics). The solid green lines represent the result obtained based on the effective medium theory, so that the energy dispersion is found by numerically solving Eq. (9). On the other hand the discrete (black) symbols are obtained from a full wave calculation of the Bloch modes of Eq. (3) using the plane wave method (Appendix A). In the full wave calculation the wave vector is oriented along the $x$-direction, $\mathbf{k} = k(1,0)$. In Fig. 2 the energy is normalized to $\hbar v_P / a = 0.09 [eV]$. The agreement between the "exact" results and the effective medium theory is excellent for $ka < 2.0$ and deteriorates near the edges of the BZ (not shown here). This validates the proposed methods.

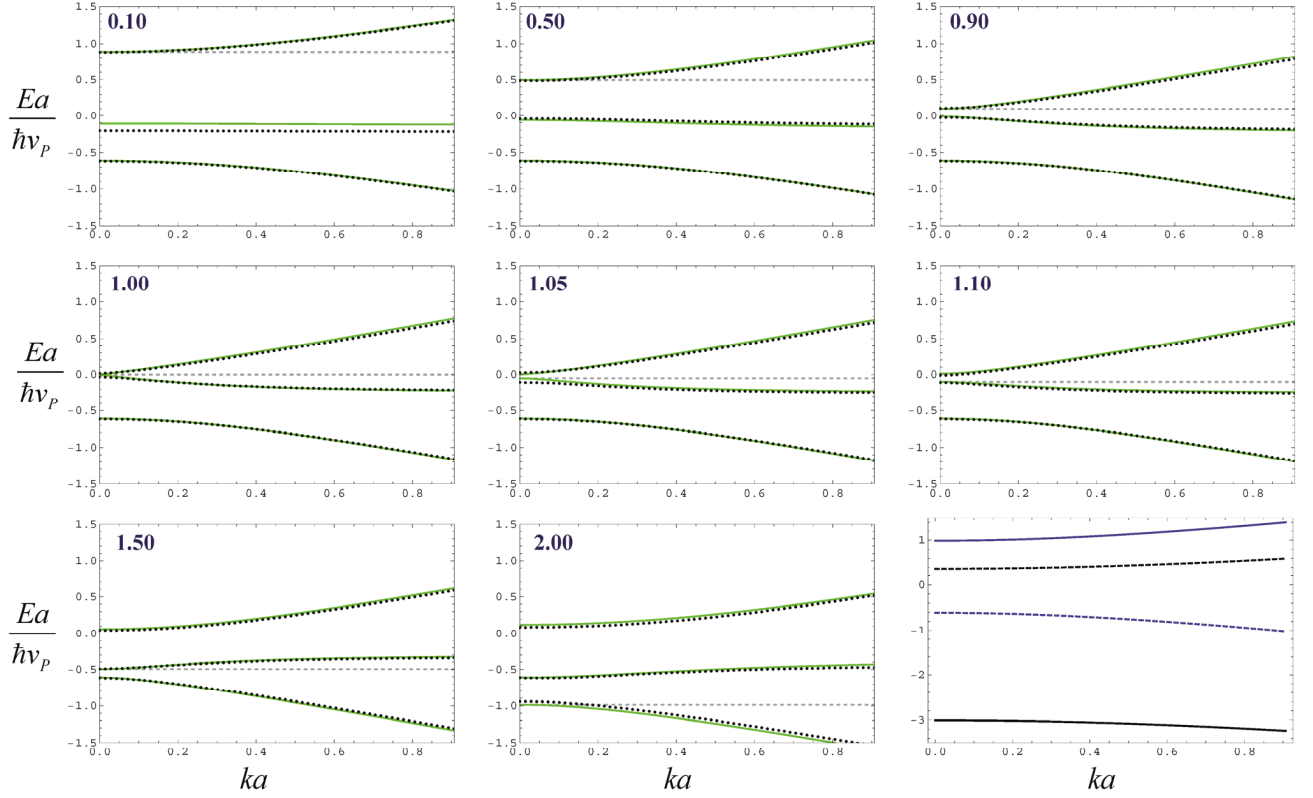

Fig. 2. Electronic band structure of a superlattice of $Hg_{0.75}Cd_{0.25}Te$ and HgTe for different values of the normalized volume fraction $f_V / f_{V0}$ of the HgTe inclusions. Solid lines: effective medium theory; Discrete points: exact band structure obtained numerically by solving Eq. (3) with the plane wave method. The HgTe inclusions have circular cross-section and are embedded in the $Hg_{0.75}Cd_{0.25}Te$ host, arranged in a triangular lattice with period $a = 12a_s$ with $a_s = 0.65nm$ being the atomic lattice constant of the bulk semiconductor alloys. The insets specify the normalized volume fraction $f_V / f_{V0}$ being $f_{V0} = 0.247$ the theoretical volume fraction for which there is a zero gap. The dashed horizontal grid line represents the energy level where $E = V_{ef}$, that is the edge of the hybridized conduction (s-type) band. The plot in the lower-right-hand side corner shows the electronic band structure of the bulk semiconductor alloys. Blue curves: $Hg_{0.75}Cd_{0.25}Te$; Black curves: HgTe. The solid curves represent the conduction ($\Gamma_6$) bands, whereas the dashed lines the valence ($\Gamma_8$) bands.

Moreover, consistent with our analytical theory, when $f_V/f_{V0}=1.0$ the electronic structure has a zero-gap and linear energy dispersion at $E=V_{ef0}$. A straightforward analysis shows that in the vicinity of $E=V_{ef0}$ the dispersive effective mass (8) has the first-order Taylor expansion:

$$m_{ef}(E) \approx \frac{1}{2v_{P,ef}^2}(E-V_{ef0}), \quad \text{with} \quad v_{P,ef} = v_P\sqrt{\frac{E_{v,h}-2V_h+E_{v,i}}{E_{v,h}-2V_h+V_i}}. \tag{15}$$

Hence, substituting this result into Eq. (9), it follows that the electronic structure in the vicinity of $E=V_{ef0}$ is determined by:

$$|E-V_{ef0}| = \hbar k v_{P,ef} \tag{16}$$

where the "effective Kane velocity" $v_{P,ef}$ plays a role analogous to the Fermi velocity in the case of graphene [23].

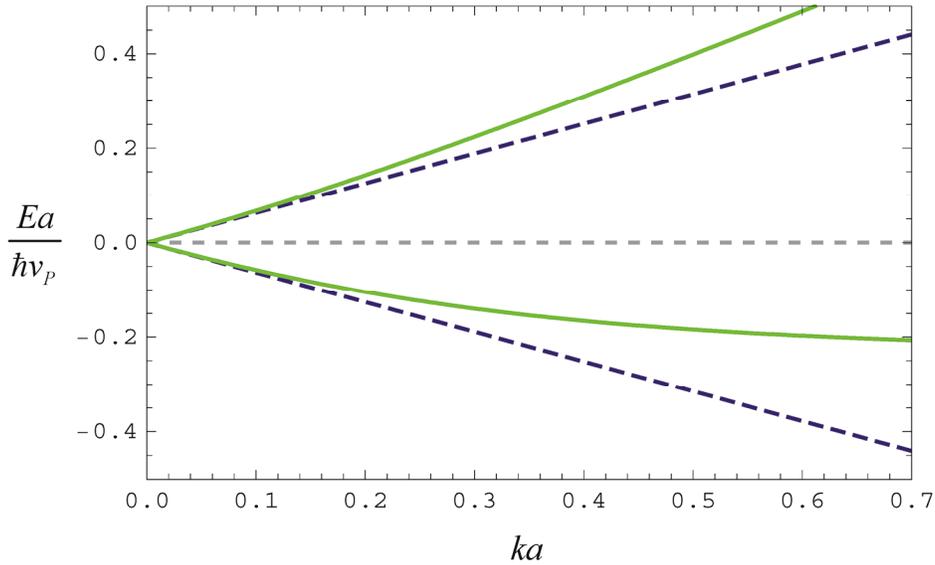

Fig. 3. Zoom of the electronic band structure of a superlattice of $Hg_{0.75}Cd_{0.25}Te$ and $HgTe$ for the case $f_V/f_{V0}=1.0$. Solid green line: analytical result obtained from Eq. (9). Dashed blue line: asymptotic

dispersion close to the contact point of the two bands [Eq. (16)]. The dashed gray horizontal grid line marks the energy level $E = V_{ef0}$.

In Fig. 3 we show a zoom of the analytical energy dispersion for the case $f_V / f_{V0} = 1.0$ close to the contact point of the two bands, superimposed on the asymptotic formula (16). For $ka < 0.1$ the asymptotic formula can be quite accurate, and the energy dispersion is practically linear. In our design $v_{P,ef} = 0.63 v_P = 6.7 \times 10^5 \, m/s$.

For values of $f_V / f_{V0}$ less than 1.0 (the first row of Fig. 2), the superlattice has a normal band structure, similar to the host material, with the conduction band above the valence bands. In Fig. 2, the edge of the conduction band is marked by the dashed horizontal grid line, and indicates the energy level for which $E = V_{ef}$. On the other hand, when $f_V / f_{V0} > 1.0$ the band gap energy is negative and the band structure is inverted, so that valence band lies above the conduction band, similar to the HgTe inclusions. Thus $f_{V0}$ plays the role of a critical volume fraction that marks the inversion of the band structure. A similar inversion of the electronic structure has been studied in the context of the spin Hall effect and occurs in HgCdTe quantum wells at a critical thickness, beyond which the transport associated with edge states becomes possible [12-14].

In Fig. 4 we represent the effective parameters ($m_{ef}$ and $E - V_{ef}$) of the superlattice for the case $f_V / f_{V0} = 1.0$. The dispersive effective mass is normalized to $\hbar / (a v_P) = 0.014 m_0$. As seen, at the energy level $E = V_{ef} = 0$ (dashed gray line) the curves of $m_{ef}$ and $E - V_{ef}$ intersect each other, confirming that the zero gap condition (10) is indeed satisfied for this superlattice. The stationary states occur for energy levels such that $m_{ef}$ and $E - V_{ef}$ have the same sign [9], which justifies that the effective medium model predicts three energy bands (Fig. 2).

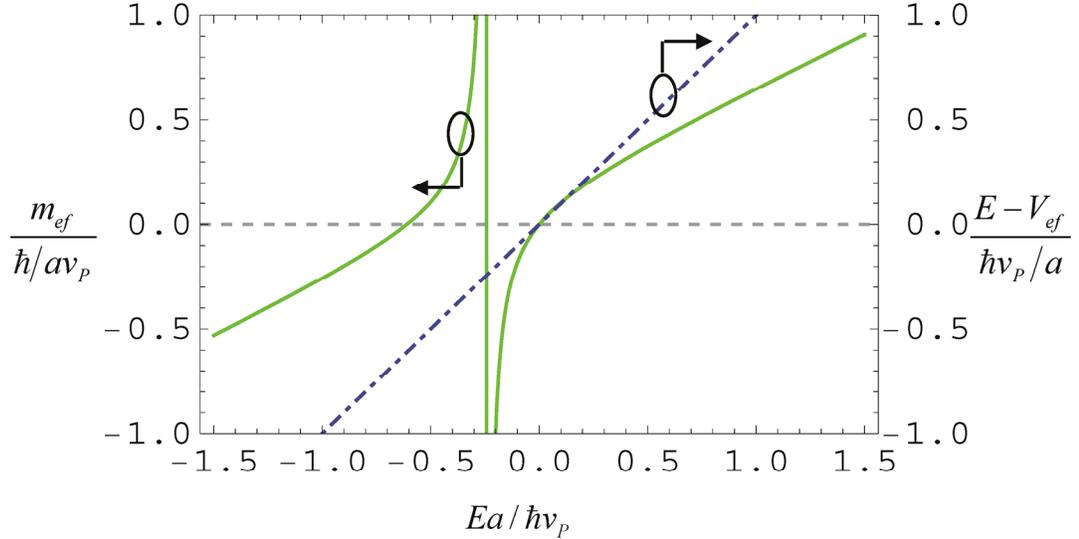

Fig. 4. Effective medium parameters of the superlattice as a function of the normalized energy $E$, for the case $f_V / f_{V0} = 1.0$. Solid green line (left axis): Normalized $m_{ef}$ [Eq. (8)]. Dashed blue line (right axis): Normalized $E - V_{ef}$ (dashed gray horizontal grid line marks the energy level $E = V_{ef}$).

As discussed in Ref. [5, 8, 9], the parameters $m_{ef}$ and $E - V_{ef}$ are in a certain sense the semiconductor analogues of the parameters $\mu$ and $\varepsilon$ in an electromagnetic metamaterial, respectively. Therefore, we see that the analogue of our zero-gap condition is $\varepsilon = 0$ and $\mu = 0$ in the electromagnetic case (matched zero-index material). Matched zero-index electromagnetic materials have been extensively studied in the literature [40-43]. The results of this section demonstrate that by controlling the geometry of the superlattice it is possible to tailor the electronic band structure, in a manner quite analogous to what is done in electromagnetic metamaterials for the photonic band structure, and that this can be accurately done using an effective medium approach.

# V. Nonlinear Electrical Response

It is known from previous studies of graphene [31-33], that a linear energy dispersion may boost the optical nonlinear response. Next, we investigate the nonlinear response of the zero-gap superlattice using the semiclassical Boltzmann theory. For simplicity, the effect of collisions is neglected and we focus the discussion in the case wherein $k_B T / \mu_c \to 0$ ("low" temperature limit), with $\mu_c$ being the chemical potential. Moreover, we consider that the inclusions are spherical "dots" so that one can assume that $|E| = \hbar k v_{P,ef}$ [see Eq. (16)] with $k = \sqrt{k_x^2 + k_y^2 + k_z^2}$ (three-dimensional configuration). We suppose that $\mu_c > 0$, and thus at low temperatures the optical response is determined by the upper energy band $E = +\hbar k v_{P,ef}$.

## A. Density of electric current in a zero-gap semiconductor

Under an external electric field excitation, $\mathbf{E}_e(t)$, the density of electric current induced in the material is [44],

$$\mathbf{j}_e(t) = -e g_s g_v \frac{1}{(2\pi)^3} \int g(\mathbf{k},t) \frac{1}{\hbar} \nabla_\mathbf{k} E d^3\mathbf{k} \qquad (17)$$

where $g_s = 2$ and $g_v = 1$ represent the spin and valley degeneracy, respectively, $E = +\hbar k v_{P,ef}$ is the dispersion of the energy states, and $g(\mathbf{k},t)$ is the non-equilibrium distribution function that determines the probability of occupation of the energy states. From Boltzmann theory, neglecting the effect of collisions, it can be proven that $g(\mathbf{k},t) = \frac{1}{1 + e^{(E(\mathbf{k}(t)) - \mu_c)/k_B T}}$, being $\mu_c$ the chemical potential and $\mathbf{k}(t) = \mathbf{k} + \delta \mathbf{k}(t)$ with $\delta \mathbf{k}(t) = \dfrac{e}{\hbar} \int\limits_{-\infty}^{t} \mathbf{E}_e(t') dt'$ [32]. Therefore, using $E = +\hbar k v_{P,ef}$ it follows that in the low temperature limit ($k_B T / \mu_c \to 0$),

$$\mathbf{j}_e(t) = -eg_s \dfrac{1}{(2\pi)^3} \int u\left(\dfrac{\mu_c}{\hbar v_{P,ef}} - |\mathbf{k}(t)|\right) \dfrac{1}{\hbar} \nabla_{\mathbf{k}} E \, d^3\mathbf{k}$$
$$= -eg_s v_{P,ef} \dfrac{1}{(2\pi)^3} \int \delta\left(\dfrac{\mu_c}{\hbar v_{P,ef}} - |\mathbf{k}(t)|\right) \dfrac{\mathbf{k}(t)}{|\mathbf{k}(t)|} k \, d^3\mathbf{k}$$
(18)

where $u$ and $\delta$ represent the Heaviside step and the Dirac delta functions, respectively, and the second identity is obtained after integration by parts. Making the change of variable $\mathbf{q} = \mathbf{k}(t)$ and using spherical coordinates $(q, \theta, \varphi)$, we obtain that

$$\mathbf{j}_e(t) = -eg_s v_{P,ef} \dfrac{1}{(2\pi)^3} \int \delta\left(q - \dfrac{\mu_c}{\hbar v_{P,ef}}\right) \hat{\mathbf{q}} |\mathbf{q} - \delta\mathbf{k}| d^3\mathbf{q}$$
$$= -eg_s v_{P,ef} \dfrac{1}{(2\pi)^3} \left(\dfrac{\mu_c}{\hbar v_{P,ef}}\right)^3 \int \hat{\mathbf{q}} |\hat{\mathbf{q}} - \mathbf{f}| d\Omega(\hat{\mathbf{q}})$$
(19)

where $\hat{\mathbf{q}} = \mathbf{q}/|\mathbf{q}|$, and $d\Omega(\hat{\mathbf{q}}) = \sin\theta d\theta d\varphi$ is the infinitesimal element of solid angle. We defined $\mathbf{f} = \dfrac{\hbar v_{P,ef}}{\mu_c} \delta \mathbf{k}$ which can be written as

$$\mathbf{f} = \dfrac{e v_{P,ef}}{\mu_c} \hat{P} \mathbf{E}_e, \qquad \text{being } \hat{P} = \int\limits_{-\infty}^{t} .\, dt'$$
(20)

an integral operator such that $\hat{P}\mathbf{E}_e = \int\limits_{-\infty}^{t} \mathbf{E}_e(t') dt'$. By choosing the polar axis aligned with $\hat{\mathbf{f}} = \mathbf{f}/|\mathbf{f}| = \hat{P}\mathbf{E}_e / |\hat{P}\mathbf{E}_e|$, it is possible to check that:

$$\int \hat{\mathbf{q}} |\hat{\mathbf{q}} - \mathbf{f}| d\Omega(\hat{\mathbf{q}}) = \hat{\mathbf{f}} \iint \cos\theta |\hat{\mathbf{q}} - \mathbf{f}| \sin\theta d\theta d\varphi$$

$$= 2\pi \hat{\mathbf{f}} \int_0^\pi \sqrt{1 + f^2 - 2f\cos\theta} \sin\theta \cos\theta d\theta, \quad (21)$$

$$= 2\pi \hat{\mathbf{f}} \left(-\frac{2}{3}f\right) G(f)$$

with $f = |\mathbf{f}| = \left|\frac{ev_{P,ef}}{\mu_c} \hat{P} \mathbf{E}_e\right|$ and

$$G(f) = \begin{cases} 1 - \dfrac{f^2}{5}, & |f| < 1 \\ \dfrac{1}{|f|} - \dfrac{1}{5}\dfrac{1}{|f|^3}, & |f| > 1 \end{cases}. \quad (22)$$

Substituting this result into Eq. (19), it follows that:

$$\mathbf{j}_e(t) = G(f) \hat{\sigma}_l \cdot \mathbf{E}_e, \qquad \text{with } \hat{\sigma}_l = \frac{e^2}{\hbar^2} g_s \frac{1}{(2\pi)^2} \frac{2}{3} \frac{\mu_c^2}{\hbar v_{P,ef}} \hat{P} \quad (23)$$

with $\hat{P}$ being the time-integration operator. In case of weak fields, $f \approx 0$ and $G(f) \approx 1$. Thus, $\hat{\sigma}_l$ represents the electrical conductivity operator in the linear regime. Note that for time harmonic fields $\hat{P} \leftrightarrow \dfrac{1}{-i\omega}$ and hence $\hat{\sigma}_l = \dfrac{ie^2}{\hbar^2 \omega} g_s \dfrac{1}{(2\pi)^2} \dfrac{2}{3} \dfrac{\mu_c^2}{\hbar v_{P,ef}}$, with $\omega$ being the frequency of oscillation of the radiation field.

Noting that the electron concentration in the upper band is $n_V = g_s \dfrac{1}{V_{tot}} \sum_{\mathbf{k}} g^0(E_{\mathbf{k}})$ ($g^0$ stands for the Fermi-Dirac distribution), which at low temperatures is

$$n_V = g_s \frac{1}{(2\pi)^2} \frac{2}{3} \left(\frac{\mu_c}{\hbar v_{P,ef}}\right)^3, \quad (24)$$

it is possible to write $\hat{\sigma}_l = e^2 n_V \dfrac{v_{P,ef}^2}{\mu_c} \hat{P}$. Hence Eq. (23) is equivalent to:

$$\mathbf{j}_e(t) = ev_{P,ef} n_V G(f)\mathbf{f}, \qquad \text{with} \quad \mathbf{f} = \frac{ev_{P,ef}}{\mu_c}\hat{\mathbf{P}}\mathbf{E}_e. \qquad (25)$$

Clearly, the nonlinear optical response is determined by the function $G(f)$. To obtain a strong nonlinear response the parameter $f$ should be comparable or greater than unity, which in case of time harmonic fields implies that:

$$|f_\omega| = \left|\frac{ev_{P,ef}}{\mu_c \omega} E_e\right| \gtrsim 1 \qquad (26)$$

A related result was obtained in Ref. [31-32], but for graphene. Very interestingly, for $|f|<1$ the function $G(f)$ depends exclusively on $f^2$, implying a pure $\chi^{(3)}$ (Kerr-type) dielectric response.

On the other hand, for $|f|\gg 1$ we find that $G(f) \approx \frac{1}{|f|}$ and hence, in such a regime, using Eq. (25) it is possible to write the electric current density as follows:

$$\mathbf{j}_e \approx \frac{1}{|f|}\hat{\sigma}_l \cdot \mathbf{E} = ev_{P,ef} n_V \frac{\hat{\mathbf{P}}\cdot\mathbf{E}}{|\hat{\mathbf{P}}\cdot\mathbf{E}|}, \qquad |f|\gg 1 \qquad (27)$$

This result can also be derived starting directly with the semiclassical equations of motion, which for an electron in the zero-gap superlattice are [44]:

$$\mathbf{v} = \frac{d\mathbf{r}}{dt} = \frac{1}{\hbar}\nabla_\mathbf{k} E = v_{P,ef}\frac{\mathbf{k}}{|\mathbf{k}|} \qquad ; \qquad \hbar\frac{d}{dt}\mathbf{k} = -e\mathbf{E}_e. \qquad (28)$$

Hence, because for a strong electric field $\mathbf{k}(t) = \mathbf{k}_{t=-\infty} - \frac{e}{\hbar}\hat{\mathbf{P}}\cdot\mathbf{E} \approx -\frac{e}{\hbar}\hat{\mathbf{P}}\cdot\mathbf{E}$, it follows that the velocity of an electron satisfies $\mathbf{v} \approx -v_{P,ef}\frac{\hat{\mathbf{P}}\cdot\mathbf{E}}{|\hat{\mathbf{P}}\cdot\mathbf{E}|}$. Using $\mathbf{j}_e = -en_V \mathbf{v}$ one recovers Eq. (27).

Equation (27) implies that in case of a sufficiently strong field the amplitude of the induced current in the material is constant, whereas the direction of the current flow is determined by the direction of $\hat{\mathbf{P}} \cdot \mathbf{E}$. In particular, for a time-harmonic electric field the current is expected to vary in time approximately as a step-like function. This is analogous to the graphene response [32]. This regime is further studied in the next subsection with a numerical example.

## *B. Numerical example*

To investigate the possibilities, we consider a zero-gap HgCdTe superlattice formed by spherical inclusions so that the energy dispersion is isotropic near the Γ point. Using Eqs. (6)-(7) in the zero-gap condition (10), it is readily found that the critical volume fraction is now

$$f_{V0} = \frac{E_{v,h} + 2E_{v,i} - 3V_h}{E_{v,h} - E_{v,i} - 3(V_h - V_i)}, \qquad \text{(spherical inclusions).} \qquad (29)$$

The effective electron velocity near the tip of the Dirac-type cone is:

$$v_{P,ef} = v_P \sqrt{\frac{E_{v,h} - 3V_h + 2E_{v,i}}{E_{v,h} - 3V_h + 2V_i}}, \qquad \text{(spherical inclusions).} \qquad (30)$$

For the same material combination as in Sect. IV, the critical volume fraction is found to be $f_{V0} = 0.219$ and the corresponding effective electron velocity is $v_{P,ef} = 0.54 v_P$. The energy dispersion can be assumed nearly linear in the range $|E - V_{ef0}| a / \hbar \tilde{v}_P < 0.1$, or equivalently for $|E - V_{ef0}| < 9\, meV$. We take the energy reference level such that $V_{ef0} = 0$. Let us suppose that the external field is time-harmonic, $\mathbf{E}_e = E_e \cos(\omega t)\hat{\mathbf{x}}$, so that, from Eq. (25) $\mathbf{j}_e(t) = e v_{P,ef} n_V G(f) f \hat{\mathbf{x}}$ with $f = f_\omega \sin(\omega t)$ and $f_\omega = \frac{e v_{P,ef}}{\mu_c \omega} E_e$. In Fig. 5a we depict the normalized $\mathbf{j}_e$ as a function of time for different values of $f_\omega$. Clearly, for $f_\omega > 1$ the response is strongly nonlinear and $\mathbf{j}_e(t)$ approaches a step-like function when $f_\omega \gg 1$.

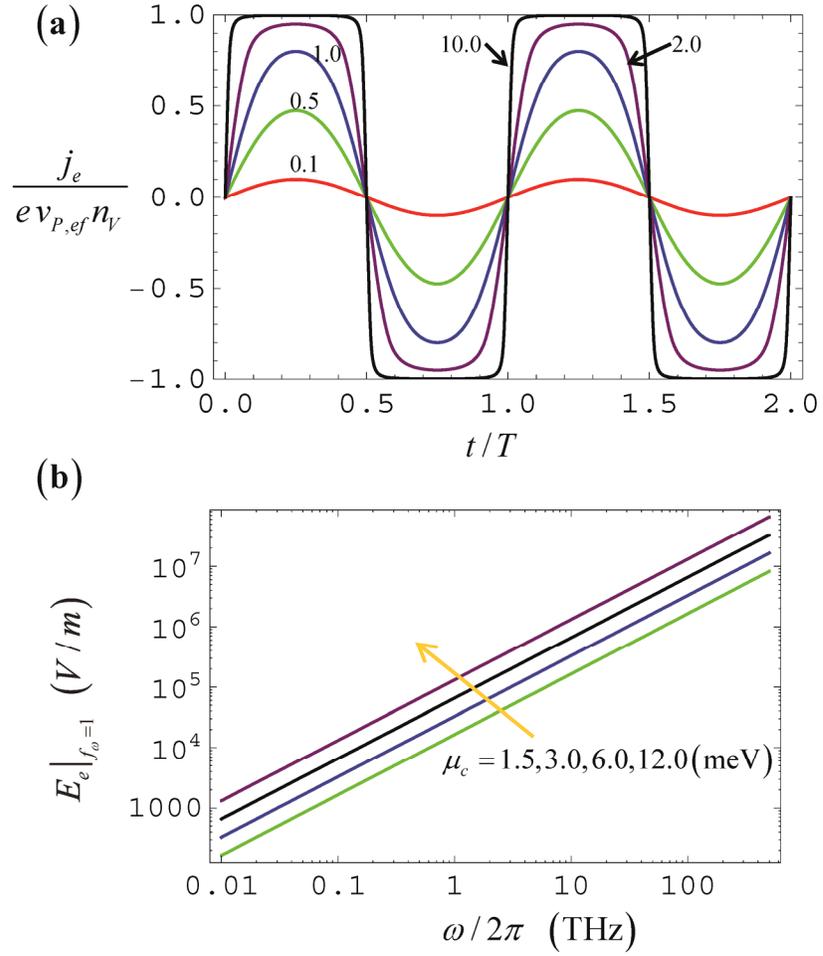

Fig. 5. (a) Normalized density of current induced in the semiconductor superlattice as a function of time (normalized to the period $T$ of the oscillation) for different values of $f_\omega = \dfrac{e v_{P,ef}}{\mu_c \omega} E_e$ shown as insets. (b) Electric field amplitude required to have a strong nonlinear response ($f_\omega = 1$) as a function of frequency for values of the chemical potential $\mu_c = 1.5,\, 3.0,\, 6.0,\, 12\ meV$ (the arrow indicates the direction of growing $\mu_c$).

In Fig. 5b we show the required electric field amplitude to have $f_\omega = 1$ as a function of frequency and for different values of the chemical potential. Note that our analysis neglects the effect of interband transitions which can play some role at terahertz frequencies and higher. As seen, stronger field amplitudes are required for larger values of $\omega$ and $\mu_c$. The dependence on $\omega$ is justified by the fact that the energy transferred by the electric field to the electron is smaller in case of a shorter period of oscillation. On the other hand, the dependence on $\mu_c$ is justified by the fact that the energy of electrons at the Fermi level is larger in case of larger $\mu_c$, and thus the relative energy change due to the applied field is smaller in case of larger values of $\mu_c$. Our results indicate that a strong nonlinear response can be obtained in the terahertz regime, extending also to the infrared and optical domains, with field amplitudes several orders of magnitude smaller than those typically required at optics. Indeed, the response in Eq. (23) leads to an equivalent complex permittivity of the form $\varepsilon \approx \varepsilon_l \langle G(f) \rangle_T = \varepsilon_l \left(1 - \langle f^2 \rangle_T / 5\right) = \varepsilon_l \left(1 - \chi |E_e|^2\right)$ for weak time-harmonic fields, where $\langle \; \rangle_T$ represents time-averaging in one period of oscillation, $\varepsilon_l$ is the equivalent permittivity response in the linear regime (which includes the effect of the conduction currents; the polarization currents due to bound charges are neglected), and

$$\chi = \frac{1}{2}\frac{f_\omega^2}{5|E_e|^2} = \frac{1}{10}\left(\frac{ev_{P,ef}}{\mu_c \omega}\right)^2. \tag{31}$$

This $\chi$ is the analogue of the $\chi^{(3)}$-parameter that characterizes the (Kerr) non-linear response of natural materials. Note that the real part of $\varepsilon_l$ is negative for low frequencies, and thus an increase of the external field results in an increase of the equivalent permittivity, similar to conventional self-focusing (but transparent) dielectric media. For $\mu_c = 12\ meV$ and $v_{P,ef} = 0.54 v_P$, we obtain $\chi = 5.8 \times 10^{-12} / f_{THz}^2\ [V^{-2} m^2]$, with $f_{THz}$ being the frequency in the THz. In the mid-infrared domain this non-linearity is several orders of magnitude larger than the $\chi^{(3)}$ of natural crystals, glasses, polymers and liquids which typically ranges in between $10^{-21}\ [V^{-2} m^2]$ and $10^{-18}\ [V^{-2} m^2]$ Ref. [45, p. 212], and is comparable to what can be achieved with semiconductor quantum wells [29] which are discrete structures, while here we have a bulk material.

# VI. Conclusion

Using the envelope function approach, we investigated semiconductor superlattices based on HgCdTe alloys, proving that the stationary energy states of the superlattice can be accurately predicted with an effective medium theory rooted in a parallelism with electromagnetics and Clausius-Mossotti formula. In particular, we have shown that for some critical volume fraction of the HgTe inclusions the band gap closes, and the superlattice is characterized by a linear energy-momentum dispersion, similar to graphene but for three-dimensional bulk systems. We studied the transport properties of the superlattice under an external electromagnetic field excitation. Based on a semiclassical approximation, we obtained an analytical formula that relates the induced current density with the external field. Our findings indicate that the response of a zero-gap superlattice may be highly nonlinear in the terahertz regime, particularly when the chemical potential is near the tip of the Dirac-type point.

**Acknowledgments:**

This work is supported in part by the US Air Force Office of Scientific Research (AFOSR) Multidisciplinary University Research Initiatives (MURI) on Quantum Metaphotonics & Metamaterials, award No. FA9550-12-1-0488), and by Fundação para a Ciência e a Tecnologia grant number PTDC/EEI-TEL/2764/2012.

## Appendix A: The electronic band structure

Here, we outline how the exact electronic band structure can be determined using the plane wave method [46]. To this end, let us consider a superlattice described by the potential $V = V(\mathbf{r})$ and by the dispersive mass $m = m(\mathbf{r}, E)$. The energy eigenstates satisfy Eq. (3). We are interested in the case where $m(\mathbf{r}, E) = \dfrac{E - E_v(\mathbf{r})}{2v_P^2} \equiv \hbar^2 A(\mathbf{r}) E - \hbar^2 B(\mathbf{r})$, with $A(\mathbf{r}) = \dfrac{1}{2\hbar^2 v_P^2}$ and $B(\mathbf{r}) = \dfrac{E_v(\mathbf{r})}{2\hbar^2 v_P^2}$. It is supposed that the geometry is two-dimensional so that the medium parameters and the (envelope) wavefunction $\Psi$ are independent of the coordinate $z$.

To determine the eigenvalues $E$ – similar to the ideas considered in Refs. [47-48] – it is convenient to reformulate the mathematical problem as a generalized eigensystem of the form $\mathbf{M} \cdot \mathbf{x} = E \, \mathbf{N} \cdot \mathbf{x}$, such that the operators $\mathbf{M}$ and $\mathbf{N}$ are independent of $E$. To this end, we introduce the auxiliary field $\mathbf{f} = (f_x, f_y, 0) \equiv \dfrac{\hbar^2}{2m} \nabla \Psi \times \hat{\mathbf{z}}$, so that from Eq. (3) $\nabla \times \mathbf{f} = -\dfrac{1}{2} \nabla \cdot \dfrac{\hbar^2}{2m} \nabla \Psi \hat{\mathbf{z}} = (E - V) \Psi \hat{\mathbf{z}}$. Thus, $(\Psi, f_x, f_y)$ satisfy the following system:

$$\hat{\mathbf{z}} \cdot \nabla \times \mathbf{f} = (E - V) \Psi, \qquad \nabla \Psi \times \hat{\mathbf{z}} = 2 \dfrac{m}{\hbar^2} \mathbf{f}. \tag{A1}$$

Using $m/\hbar^2 = AE - B$, this can also be written in a matrix form,

$$\begin{pmatrix} V & -\dfrac{\partial}{\partial y} & \dfrac{\partial}{\partial x} \\ \dfrac{\partial}{\partial y} & 2B & 0 \\ -\dfrac{\partial}{\partial x} & 0 & 2B \end{pmatrix} \begin{pmatrix} \Psi \\ f_x \\ f_y \end{pmatrix} = E \begin{pmatrix} 1 & 0 & 0 \\ 0 & 2A & 0 \\ 0 & 0 & 2A \end{pmatrix} \begin{pmatrix} \Psi \\ f_x \\ f_y \end{pmatrix} \quad \text{(A2)}$$

which corresponds to the desired $\mathbf{M}\cdot\mathbf{x} = E\mathbf{N}\cdot\mathbf{x}$ structure, with $\mathbf{M}$ and $\mathbf{N}$ Hermitian operators and $\mathbf{N}$ positive definite because $A(\mathbf{r}) > 0$.

It should be noted that $V$, $A$, and $B$ depend on the position. To determine the electronic band structure $\Psi$, $f_x$ and $f_y$ are expanded into plane waves [46]:

$$\Psi = \sum_{\mathbf{J}} \Psi_{\mathbf{J}} e^{i\mathbf{k_J}\cdot\mathbf{r}}, \qquad f_i = \sum_{\mathbf{J}} f_{\mathbf{J}}^{(i)} e^{i\mathbf{k_J}\cdot\mathbf{r}}, \quad i=x,y \quad \text{(A3)}$$

where $\Psi_{\mathbf{J}}$ and $f_{\mathbf{J}}^{(i)}$ are the coefficients of the plane wave expansion, $\mathbf{k_J} = \mathbf{k} + \mathbf{k_J^0}$ with $\mathbf{k} = (k_x, k_y)$ the quasi-momentum, and $\mathbf{k_J^0}$ represents a generic point of the reciprocal lattice.

We restrict our analysis to the case in which the unit cell corresponds to a uniform host loaded with "inclusions" with circular cross section with radius $R$. Then, for $\mathbf{r}$ in the unit cell it is possible to write:

$$V(\mathbf{r}) = V_h + \delta V \chi_{inc}(\mathbf{r}) \quad \text{(A4a)}$$

$$A(\mathbf{r}) = A_h + \delta A \chi_{inc}(\mathbf{r}), \qquad B(\mathbf{r}) = B_h + \delta B \chi_{inc}(\mathbf{r}) \quad \text{(A4b)}$$

where $\chi_{inc}(\mathbf{r}) = 0$ in the host region and $\chi_{inc}(\mathbf{r}) = 1$ in the cylindrical inclusion, so that $V = V_h$ in the host region, and $V = V_h + \delta V$ in the inclusion, etc. The characteristic function $\chi_{inc}(\mathbf{r})$ can be written in terms of a plane wave expansion as follows:

$$\chi_{inc}(\mathbf{r}) = \sum_{\mathbf{J}} \chi_{\mathbf{J}} e^{i\mathbf{k}_{\mathbf{J}}^0 \cdot \mathbf{r}}, \text{ with } \chi_{\mathbf{J}} = \frac{\pi R^2}{A_c} \frac{2J_1(|\mathbf{k}_{\mathbf{J}}^0|R)}{|\mathbf{k}_{\mathbf{J}}^0|R} \quad (A5)$$

where $R$ is the radius of the inclusion and $A_c$ is the area of the unit cell. Thus,

$$V(\mathbf{r}) = \sum_{\mathbf{J}} V_{\mathbf{J}} e^{i\mathbf{k}_{\mathbf{J}}^0 \cdot \mathbf{r}} \text{ with } V_{\mathbf{J}} = V_h \delta_{\mathbf{J},0} + \delta V \chi_{\mathbf{J}}, \quad A(\mathbf{r}) = \sum_{\mathbf{J}} A_{\mathbf{J}} e^{i\mathbf{k}_{\mathbf{J}}^0 \cdot \mathbf{r}} \text{ with } A_{\mathbf{J}} = A_h \delta_{\mathbf{J},0} + \delta A \chi_{\mathbf{J}},$$

and $B(\mathbf{r}) = \sum_{\mathbf{J}} B_{\mathbf{J}} e^{i\mathbf{k}_{\mathbf{J}}^0 \cdot \mathbf{r}}$ with $B_{\mathbf{J}} = B_h \delta_{\mathbf{J},0} + \delta B \chi_{\mathbf{J}}$. Substituting these formulas and the plane wave expansion (A3) into the generalized eigensystem (A2), it is easily found that it can be rewritten as:

$$\begin{pmatrix} [\tilde{V}_{\mathbf{I},\mathbf{J}}] & [-\delta_{\mathbf{I},\mathbf{J}} i\mathbf{k}_\mathbf{I} \cdot \hat{\mathbf{y}}] & [\delta_{\mathbf{I},\mathbf{J}} i\mathbf{k}_\mathbf{I} \cdot \hat{\mathbf{x}}] \\ [\delta_{\mathbf{I},\mathbf{J}} i\mathbf{k}_\mathbf{I} \cdot \hat{\mathbf{y}}] & [2B_{\mathbf{I},\mathbf{J}}] & 0 \\ [-\delta_{\mathbf{I},\mathbf{J}} i\mathbf{k}_\mathbf{I} \cdot \hat{\mathbf{x}}] & 0 & [2B_{\mathbf{I},\mathbf{J}}] \end{pmatrix} \begin{pmatrix} \Psi_{\mathbf{J}} \\ f_{\mathbf{J}}^{(x)} \\ f_{\mathbf{J}}^{(y)} \end{pmatrix} = E \begin{pmatrix} [\delta_{\mathbf{I},\mathbf{J}}] & 0 & 0 \\ 0 & [2A_{\mathbf{I},\mathbf{J}}] & 0 \\ 0 & 0 & [2A_{\mathbf{I},\mathbf{J}}] \end{pmatrix} \begin{pmatrix} \Psi_{\mathbf{J}} \\ f_{\mathbf{J}}^{(x)} \\ f_{\mathbf{J}}^{(y)} \end{pmatrix}$$

(A6)

where $[A_{\mathbf{I},\mathbf{J}}]$ is a square matrix such that $A_{\mathbf{I},\mathbf{J}} \equiv A_{\mathbf{I}-\mathbf{J}}$, $[B_{\mathbf{I},\mathbf{J}}]$ and $[V_{\mathbf{I},\mathbf{J}}]$ are defined similarly, $[\delta_{\mathbf{I},\mathbf{J}}]$ is an identity matrix ($\delta_{\mathbf{I},\mathbf{J}}$ represents the Kronecker's symbol), and $[\delta_{\mathbf{I},\mathbf{J}} i\mathbf{k}_\mathbf{I} \cdot \hat{\mathbf{x}}]$ represents a diagonal matrix with element $(\mathbf{I}, \mathbf{J})$ determined by $\delta_{\mathbf{I},\mathbf{J}} i\mathbf{k}_\mathbf{I} \cdot \hat{\mathbf{x}}$. The above generalized matrix eigensystem can be easily numerically solved using standard methods. Note that both matrices are Hermitian, and the matrix on the right-hand side is positive definite.

Similar to the results of Ref. [47], it turns out that the numerical solution of Eq. (A6) yields an extremely large number of flat (dispersionless, i.e. independent of $\mathbf{k}$) bands with accumulation point at the energy value where $m_h(E) = 0$. These bands are the analogue of "plasmons" in electromagnetics, and are related in the context of semiconductor superlattices to the heavy hole band. For clarity, in the numerical results of section IV, we only report the non-dispersionless bands which are typically superimposed on the flat bands. The flat bands are removed from the band structure with a suitable algorithm, whose discussion is out of the scope of the present work.

## Appendix B: The effective mass

Here, we derive the expression for the dispersive effective mass $m_{ef}$ of the superlattice. To do this, we consider that $V(\mathbf{r}, E) \approx const.$ and that the energy is such that $E \approx V$, so that the Schrödinger-type equation (3) reduces to $\nabla \cdot \left( \frac{1}{m} \nabla \Psi \right) \approx 0$. This equation is formally analogous to the equation $\nabla \cdot (\varepsilon \nabla \phi) = 0$ that occurs in electrostatics, with $\varepsilon = \varepsilon(\mathbf{r})$ being the electric permittivity and $\phi$ being the electric potential (so that $\mathbf{D} = -\varepsilon \nabla \phi$ is the electric displacement field). We are interested in the case where $\varepsilon = \varepsilon(\mathbf{r})$ is a periodic function of the spatial coordinates, so that the structure may be seen as a photonic crystal. In the electrostatic limit, the effective permittivity of a photonic crystal can be predicted using the Clausius-Mossotti formula [49, 50]

$$\varepsilon_{ef} = \varepsilon_h \left( 1 + \frac{\alpha_n}{1 - C \alpha_n} \right), \tag{B1}$$

where $\varepsilon_h$ represents the permittivity of the host region, $\alpha_n$ is the electric polarizability of the inclusion normalized to the volume of the unit cell, and $C$ is a normalized interaction constant that equals $1/3$ for lattices with high symmetry (e.g. lattices with either f.c.c., b.c.c., or simple cubic symmetry). For the case of a spherical inclusion with permittivity $\varepsilon_i$ and volume fraction $f_V$ the normalized polarizability satisfies $\alpha_n = 3f_V \dfrac{\varepsilon_i - \varepsilon_h}{\varepsilon_i + 2\varepsilon_h}$ [51, p. 139]. Hence, the effective permittivity of an array of dielectric spheres embedded in a dielectric host can be written as:

$$\varepsilon_{ef} = \varepsilon_h \frac{(1+2f_V)\varepsilon_i + 2(1-f_V)\varepsilon_h}{(1-f_V)\varepsilon_i + (2+f_V)\varepsilon_h}. \tag{B2}$$

The similarity between the equations $\nabla \cdot \left( \dfrac{1}{m} \nabla \Psi \right) = 0$ and $\nabla \cdot (\varepsilon \nabla \phi) = 0$ suggests that the effective mass $m_{ef}$ can be obtained from the following transformations $1/\varepsilon_{ef} \to m_{ef}$, $1/\varepsilon_i \to m_i$ and $1/\varepsilon_h \to m_h$, so that:

$$m_{ef} = m_h \frac{(1-f_V)/m_i + (2+f_V)/m_h}{(1+2f_V)/m_i + 2(1-f_V)/m_h}. \tag{B3}$$

After some simplifications this yields Eq. (7).

In the two-dimensional case, such that the superlattice is formed by a highly symmetric lattice of cylindrical inclusions with radius $R$, the formula for $m_{ef}$ can be obtained in the same manner. For such a geometry it is evident that $C = 1/2$ (e.g. for either a square or triangular lattice), and from [51, p. 145; the geometrical factor of Eq. 5.32 is $L = 1/2$ for a cylinder] the normalized polarizability is $\alpha_n = 2f_V \dfrac{\varepsilon_i - \varepsilon_h}{\varepsilon_i + \varepsilon_h}$. Thus, the effective permittivity is now:

$$\varepsilon_{ef} = \varepsilon_h \frac{\varepsilon_i(1+f_V) + \varepsilon_h(1-f_V)}{\varepsilon_i(1-f_V) + \varepsilon_h(1+f_V)}. \tag{B4}$$

Hence, applying again the transformation $1/\varepsilon_{ef} \to m_{ef}$, one obtains Eq. (8).